\documentclass[iop,apj]{emulateapj}
\usepackage[usenames,dvipsnames,svgnames]{xcolor}
\usepackage{epstopdf}
\usepackage{hyperref}
\usepackage{zref}
\usepackage{xr}
\newcommand{\Msun}{\ensuremath{{\rm M}_\odot}}

\newcommand{\kms}{\ensuremath{\,{\rm km\,s}^{-1}}}

\newcommand{\fgas}{\ensuremath{f_{gas}}}

\begin{document}

\title{Gas Content and Kinematics in Clumpy, Turbulent Star-forming Disks}

\author{Heidi A. White\altaffilmark{1,2}, David B. Fisher\altaffilmark{3}, Norman Murray\altaffilmark{4}, Karl Glazebrook\altaffilmark{3}, Roberto G. Abraham\altaffilmark{1,2}, Alberto D. Bolatto\altaffilmark{5}, Andrew W. Green\altaffilmark{6}, Erin Mentuch Cooper\altaffilmark{7}, Danail Obreschkow\altaffilmark{8}}
\altaffiltext{1}{Department of Astronomy and Astrophysics, University of Toronto, 50 St. George St., Toronto, ON, M5S 3H8, Canada}
\altaffiltext{2}{Dunlap Institute for Astronomy and Astrophysics, University of Toronto, 50 St. George St., Toronto, ON, M5S 3H8, Canada}
\altaffiltext{3}{Centre for Astrophysics and Supercomputing, Swinburne University of Technology, P.O. Box 218, Hawthorn, VIC 3122, Australia}
\altaffiltext{4}{Canadian Institute for Theoretical Astrophysics, 60 St. George Street, University of Toronto, Toronto ON M5S 3H8, Canada}
\altaffiltext{5}{Department of Astronomy and Joint Space Institute, University of Maryland, College Park, MD 20642, USA}
\altaffiltext{6}{Australian Astronomical Observatory, P.O. Box 970, North Ryde, NSW 1670, Australia}
\altaffiltext{7}{Astronomy Department, University of Texas at Austin, Austin, TX 78712, USA}
\altaffiltext{8}{International Centre for Radio Astronomy Research (ICRAR), University of Western Australia, M468, Crawley, WA 6009, Australia}

\begin{abstract}
We present molecular gas mass estimates for a sample of 13 local galaxies whose kinematic and star forming properties closely resemble those observed in $z\approx 1.5$ main-sequence galaxies. Plateau de Bure observations of the CO[1-0] emission line and {\em Herschel Space Observatory} observations of the dust emission both suggest molecular gas mass fractions of $\sim$20\%. Moreover, dust emission modeling finds $T_{dust}<$30K, suggesting a cold dust distribution compared to their high infrared luminosity. 
%have been performed for six galaxies (of which five are well-detected at $>$8$\sigma$) yielding CO fluxes and line luminosities consistent with gas mass fractions up to $\sim$30\%, assuming $\alpha_{CO}$=3.1 $\Msun$(K km s$^{-1}$ pc$^{2}$)$^{-1}$.
%Fitting a modified blackbody function to existing Herschel IR observations (from PACS+SPIRE) for ten additional galaxies, we find substantial dust masses ($M_{dust}$; 1-3$\times 10^{9}\Msun$) and $T_{dust}<$30K, suggesting that the dust within these systems is cold.Application of the locally-derived dust-to-gas ratio (D:G$\sim$0.01) to fitted $M_{dust}$ suggests $f_{gas}$ $\sim$10-40\%.
The gas mass estimates argue that $z\sim$0.1 DYNAMO galaxies not only share similar kinematic properties with high-z disks, but they are also similarly rich in molecular material. Pairing the gas mass fractions with existing kinematics reveals a linear relationship between $f_{gas}$ and $\sigma$/$v_{c}$, consistent with predictions from stability theory of a self-gravitating disk. It thus follows that high gas velocity dispersions are a natural consequence of large gas fractions. We also find that systems with lowest $t_{dep}$ ($\sim$0.5 Gyr) have the highest ratios of $\sigma$/$v_{c}$ and more pronounced clumps, even at the same high molecular gas fraction.
\end{abstract}

%-------------------------------------------------------
%                      INTRO
%-------------------------------------------------------
\section{Introduction}\label{sec:intro}
Observations show that the majority of star formation in the universe occurs between 1$<z<$3 \citep{hopkins2006,madau2014}. When observed in the near infrared (NIR), star forming galaxies (SFGs) at this epoch are frequently irregular or ``clumpy" in morphology and H$\alpha$ fluxes suggest elevated star formation rates (SFRs; typically $\sim$10$^{2} \Msun$ yr$^{-1}$) reminiscent of local merging systems. Integral-field spectroscopy (IFS) surveys of rest-frame optical emission lines report that despite the morphology, the kinematics of these galaxies better resemble rotating, but turbulent, disks: a large fraction of these systems exhibit ordered rotation and are observed to sit on the star-forming main sequence. Spatially-resolved observations reveal that the bulk of these systems have high internal velocity dispersions ($\sigma\approx20-100$ km/s) when compared to z$\sim$0 galaxies (see \citealt{fs2009}; \citealt{genzel2008}; \citealt{wisnioski2011,wisnioski2015}). Moreover, these systems are distinct from local star forming spirals in that they are consistently observed to contain substantial molecular gas fractions ($f_{gas}\sim$20-50\%; \citealt{tacconi2010,tacconi2013}; \citealt{daddi2010}).

A significant fraction of star formation within high-redshift clumpy galaxies occurs in large, massive clumps (kpc-scale, $\sim$10$^{9}\ \Msun$). \cite{fisher2017letter} shows that the detailed properties of clumps in turbulent disks are best described by predictions from self-gravitating instabilities within disk galaxies (as opposed to mergers or other instabilities). This theory (see \citealt{dekel2009}) suggests that the amplitude of instabilities is governed by three forces: (1) gravitational forces scaling with the surface density of the gas, (2) turbulent pressure forces due to the velocity dispersion and (3) shear forces caused by the differential rotation of the disk. The balance between these forces is suitably measured by the non-dimensional parameter Q \citep{toomre1964}, such that Q$>$1 regions are stable, whereas Q$<$1 regions are unstable and form clumps. According to this model, the high gas fractions observed in high-z disks (e.g. \citealt{elmegreen2009}) are, at least partially, responsible for the apparent widespread instabilities of these high-z disks. In turn, the high velocity dispersions (a stabilizing force) are commonly used to explain the large size of the clumps following Jeans theory. Moreover, if this mechanism is indeed the primary force in producing clump formation, it's certain to play an important role in the feedback cycle within galaxies: e.g. the release of gravitational potential energy as massive clumps form, torques felt between in-spiraling clumps, and energy injection from star formation are all likely to contribute to high velocity dispersions of the ISM (\citealt{bournaud&elmegreen2009}; \citealt{lehnert2009}; \citealt{genzel2008,genzel2011}).

Although the groundwork for this theory has existed for some time (\citealt{dekel2009}), it has just begun to be tested observationally. Truly robust tests of this instability argument requires observations of the internal properties of a sample of galaxies - gas mass fractions and resolved kinematics. Direct observation presents substantial challenges as these clumpy star forming systems are almost entirely unique to the high-z universe and are, thus, quite difficult to observe. Resolving the kinematics of the star forming and molecular regions within $1<z<3$ disks is challenging due to observational limitations: e.g. seeing/atmospheric effects, low signal-to-noise ratio (SNR). Both NIR and molecular gas observations at high-z require long integration times to ensure detection and, at present, while there exist $\sim$200 galaxies with gas fractions at high-z, only a handful have measured resolved kinematics (primarily those with overlap in the PHIBSS \& SINS samples). These high-z galaxies are also chosen to be very bright, necessarily biasing high-z observations.

An increasing number of studies focus on overcoming issues of distance by identifying rare, nearby galaxies with properties similar to high redshift main-sequence galaxies. Other groups have also identified large gas fraction systems using atomic gas \citep{garland2015,catinellacortese2015}. \cite{green2014} presented DYNAMO, a sample of 95 local ($z\sim 0.06-0.08$ \& $0.12-0.16$) galaxies whose kinematic and star formation properties closely resemble that observed in high-z clumpy disks (see \citealt{green2014}, \citealt{bassett2014}, \citealt{fisher2017}). Green et al. use an inverted Kennicutt-Schmidt relation to estimate the total gas content ($f_{gas,tot}$) for galaxies in DYNAMO and find evidence of a correlation with $\sigma$/$v_{c}$. Confirmed detections of CO[1-0] emission in four DYNAMO targets by \citet{fisher2014} argues that at least some fraction of the sample are also gas rich ($f_{gas}\sim$20-30\%).

In this paper, we analyze the ISM properties of clumpy, turbulent disk galaxies.  We present molecular gas fractions for 13 DYNAMO galaxies. We utilize two, separate methods for inferring $f_{gas}$ to limit methodological bias and maximize the sample size. We then couple this new information with existing high-resolution integral field spectroscopy (IFS) to investigate the Toomre instability argument and quantify the relationship between the molecular gas content and ionized gas kinematics. 

The paper is structured as follows: in \S\ref{sec:2}, we provide a comparison of our DYNAMO sample with systems observed at high-$z$. In \S\ref{sec:3}, we discuss our IR and CO[1-0] observations and describe the two methods utilized in estimating molecular gas and dust masses for our sample. Finally, in \S\ref{sec:4}, we present our results and discuss them in context, and in \S\ref{sec:5}, summarize our major findings.

%-------------------------------------------------------
%   DYNAMO GALAXIES ARE CLUMPY AND TURBULENT
%-------------------------------------------------------

\section{Sample}\label{sec:2}
\begin{figure}[hbt]
\begin{center}
\includegraphics[scale=0.575]{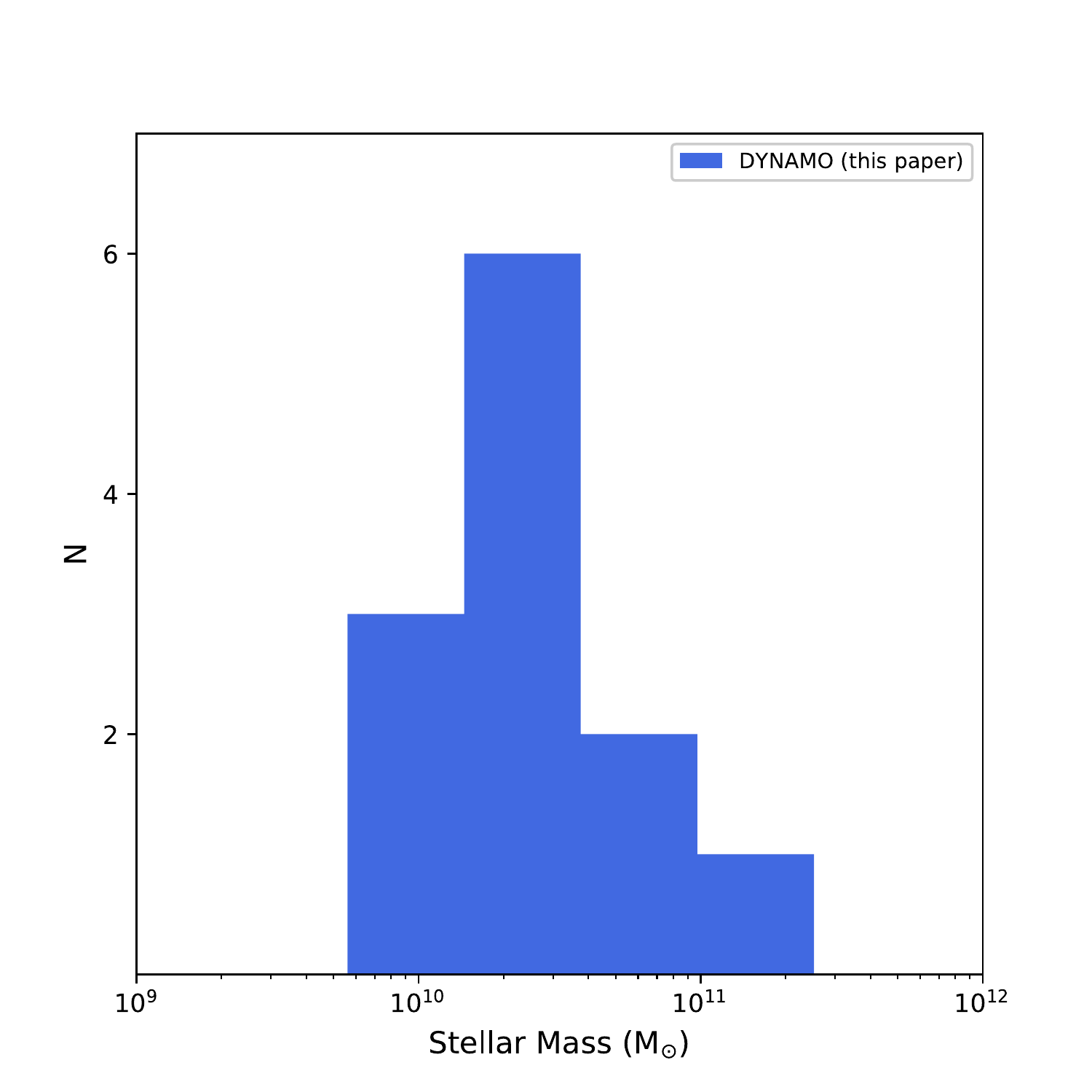}
\end{center}
\caption{The distribution of stellar masses for the DYNAMO galaxies presented in this paper. The majority of the targets within our sample range between masses of $1-5\times 10^{10} M_{\odot}$.}
\label{fig:masshist}
\end{figure}

The targets in this paper are a subset of the greater DYNAMO sample \citep[originally presented in ][hereafter referred to as DYNAMO-I]{green2014}. DYNAMO is an H$\alpha$ IFU survey of local (z$\sim$0.1) galaxies which have been selected (in two SDSS redshift windows) to be H$\alpha$-luminous (top 1\% in the local universe, based on fiber luminosity; $\overline{SFR}\sim$ 11 $\Msun$ yr$^{-1}$). The majority of stellar masses of our sample fall between 1-5$\times 10^{10}$ $\Msun$ (see Fig.\ref{fig:masshist}).

\subsection{DYNAMO systems resemble z=1.5 main-sequence galaxies}
A large fraction ($\sim$84\%) of DYNAMO galaxies appear disk-like and about half are located on the Tully-Fisher relation \citep{green2014}. Spatially resolved spectroscopy of redshifted H$\alpha$ emission in DYNAMO-I reveals enhanced velocity dispersions (20-100 km s$^{-1}$; \citealt{green2014}, \citealt{bassett2014}, \citealt{bekiaris2016}) and evidence of ordered, rotating disk structure. \cite{bassett2014} follow up on these findings by \citet{green2010,green2014} with higher resolution and more sensitive kinematics from Gemini/GMOS to confirm DYNAMO galaxies are rotating in both gas and stars.

DYNAMO targets also appear gas rich. \cite{fisher2014} performed CO[1-0] observations using the Plateau de Bure interferometer (PdBI) for four DYNAMO targets (with three successful detections) and report gas mass fractions between 20-30\%. Comparison with typical local spirals of similar stellar mass ($f_{gas}\sim$1-8\%; \citealt{saintonge2011a}) suggests that the DYNAMO galaxies comprise a very gas rich subset of the local star forming population.

Similar to systems at high-$z$ (e.g. \citealt{genzel2011,wisnioski2015}) star forming disks in DYNAMO appear not only turbulent, but clumpy. High-resolution ($\sim$100pc) Hubble Space Telescope (HST) follow up of the clumps by \cite{fisher2017} confirm that they are both large ($\overline{d_{\rm core}}\sim$0.5 kpc) and massive ($\sim$10$^{8-9}\Msun$). Moreover, adaptive optics (AO)-corrected IFS observations verify that the observed high $\sigma$ values are consistent down to sub-clump ($\sim$100-200 pc) scales (Oliva-Altamirano et al., \textit{submitted}).

\cite{fisher2017} find that DYNAMO disks are compact and their H$\alpha$ half-light radii are most similar to those observed at $1.5\leq z \leq2$. While in local systems, typical SFR surface densities ($\Sigma_{\rm SFR}$) range from $10^{-4}$-$10^{-1}$ $\rm \Msun\ yr^{-1}\ kpc^{-2}$, \cite{fisher2017} also show that $\Sigma_{\rm SFR}$ in DYNAMO is orders of magnitude greater ($\sim$ 1-10 $\rm \Msun\ yr^{-1}\ kpc^{-2}$) and best matched to that found in systems at $1<z<3$.

A plausible interpretation of DYNAMO galaxies is that their striking similarity with $z \sim 1.5$ main-sequence disks is because a similar internal physical mechanism is driving the star formation and kinematics in both objects. Indeed, \cite{green2014} and \cite{fisher2017} find that DYNAMO galaxies lie close to the $z\sim 1.5-2.0$ star formation rate-stellar mass main sequence. The key difference between these two populations, however, is not the internal physics, but rather the frequency of that mechanism at $z>1$ and at $z=0.1$. DYNAMO galaxies are then excellent laboratories for studying a major mode of high-z star formation (at a critical epoch) on higher signal-to-noise data.

\subsection{Classifying DYNAMO galaxies as mergers or disks}
Constraining the fraction of mergers within our sample has important implications for our choice of $\alpha_{\rm CO}$ and our interpretation of these DYNAMO galaxies as clumpy, turbulent disks. All of the targets in this paper have been observed to have kinematic signatures that best resemble disks (i.e. rotation and a somewhat uniform velocity dispersion field; see \citealt{green2014} and \citealt{bekiaris2016}). As we mention above, \cite{green2014} show that the kinematics in DYNAMO galaxies are consistent with the Tully-Fisher relationship.

\cite{fisher2017} use HST 600nm continuum maps to show that the surface brightness profiles of 8 of 10 DYNAMO galaxies in their sample are well described by an exponential decay with increasing radius, which is consistent with disks. Seven of these galaxies make up over half of the sample discussed in this paper (G14-1, D15-3, D13-5, C13-1, G20-2, G08-5, G04-1).

Classification of disks and mergers in galaxies with as high gas fractions and star formation rates as DYNAMO is of course an imperfect process. Moreover, the classification of ``turbulent disks" is relatively new. Nonetheless, we expect (based on our previous results) our classifications to be accurate at the $\sim$80\% level.

%-------------------------------------------------------
%            GAS CONTENT & KINEMATICS
%-------------------------------------------------------

\section{Gas content and kinematics} \label{sec:3}

\subsection{\texorpdfstring{$M_{gas}$ Determinations}{Mgas Determinations}} \label{sec:Mgas}
There are significant challenges to directly observing molecular gas within galaxies. The H$_{2}$ molecule, while abundant, has no dipole moment and its lowest vibrational state is difficult to excite at typical molecular cloud temperatures  \citep[10-20K;][]{kennicutt2012}. Consequently, inferring a galaxy's molecular gas content is most often performed via indirect measurements. In this paper, we use two common methods:

\begin{itemize}
\item Observations of ground state CO[1-0] rotational line emission allow for determination of a CO line luminosity for the galaxy. Then, using locally-derived empirical values for $\alpha_{CO}$, this CO luminosity is converted to an estimate for the system's molecular gas mass ($M_{mol}$) and, subsequently, the inferred baryonic gas fraction ($f_{gas}$ = $M_{mol}$/($M_{mol}$ + $M_{star}$)). 
\item Available IR-band flux information from Herschel PACS+SPIRE is fit with a modified blackbody model to constrain the dust mass ($M_{dust}$), which is then used to infer a gas mass via the metallicity-dependent empirical dust-to-gas ratio.
\end{itemize}

%-------------------------------------------------------
%                 CO[1-0] OBSERVATIONS
%-------------------------------------------------------

\subsubsection{CO(1-0) Observations}\label{sec:co_obs}
\begin{figure*}[!htbp]
\begin{center}
\includegraphics[scale=0.375]{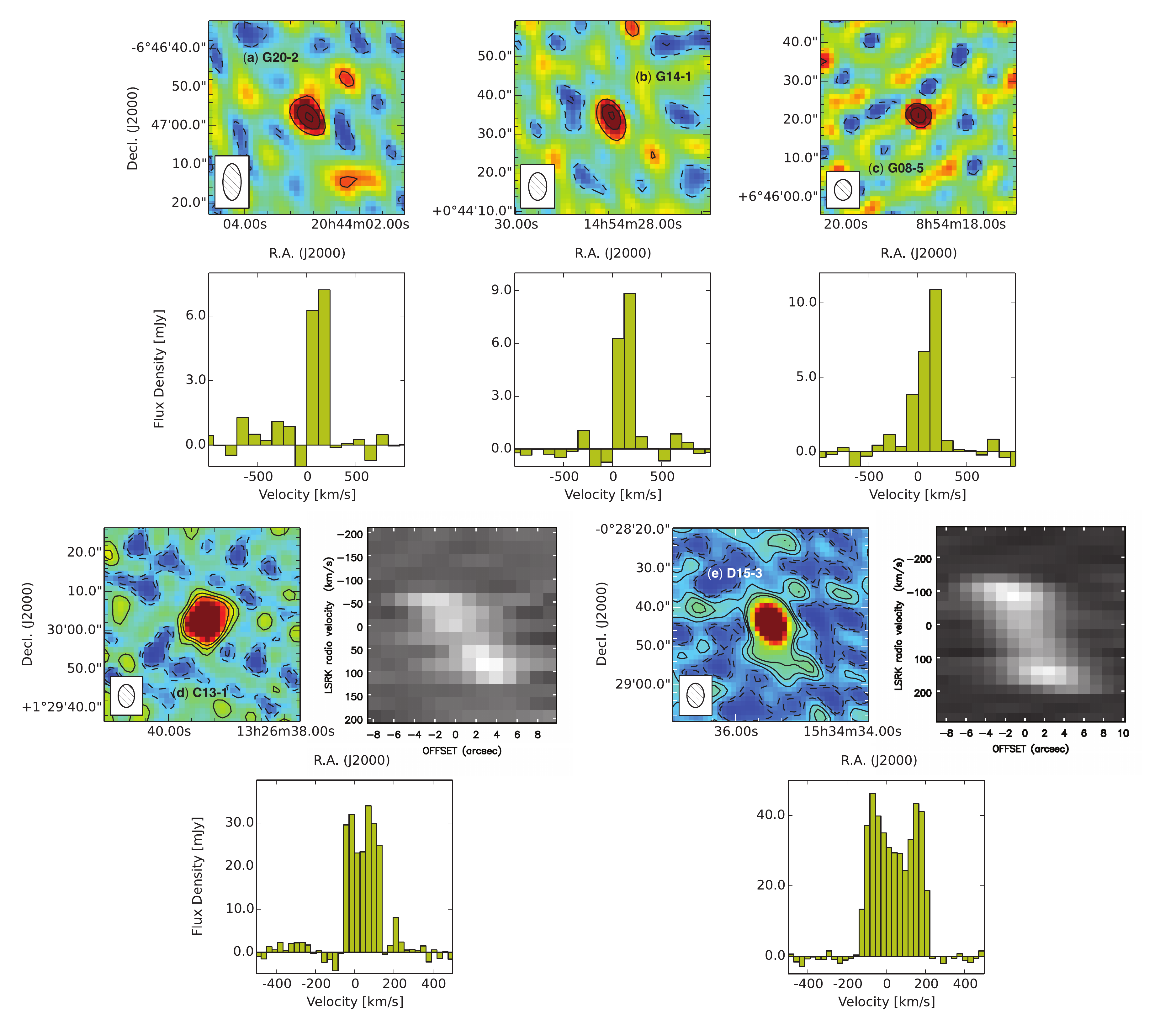}
\end{center}
\caption{Emission maps (rows 1 \& 3) and spectra (rows 2 \& 4) for the five galaxies detected with PdBI. All emission maps are 50$\arcsec \times$50$\arcsec$ and produced by collapsing across the CO[1-0] linewidth; 2-, 4-, \& 6-$\sigma$ contours and individual synthesized beams have been included for reference. Below each emission map, we include corresponding spectra of the detected CO[1-0] line referenced to the expected redshift velocity. Spectra have bin-sizes of 119 km s$^{-1}$ for galaxies G20-2 (a), G14-1 (b), \& G08-5 (c) and 28 km s$^{-1}$ for galaxies C13-1 (d) \& D15-3 (e). The two z$\sim$0.07 systems (d \& e) are marginally resolved by the beam and exhibit clear kinematic structure both on-sky and in the spectra. For these two galaxies, we have included position-velocity diagrams in row 3.}
\label{fig:co_spec}
\end{figure*}

Six galaxies (C13-1, G20-2, G13-1, G14-1, G08-5, and D15-3 from DYNAMO-I) were observed on May 20, 23, \& 27-29 of 2014 (PID 12977, PI Damjanov) for a combined period of 7.5 hr using the PdBI targeting emission within the 3-mm atmospheric window (80 - 116 GHz) via the CO(J=1$\rightarrow$0) transition. Observations were carried out using 5 antennas (in `D' configuration) and resulted in synthesized beams around 6$\arcsec$x4.5$\arcsec$ in size. Due to the declination of some of our sources and incomplete (u,v) coverage, some of our maps exhibit elongated beams and side-lobe structure (see Fig. \ref{fig:co_spec}). However, this does not significantly affect the measurement of total flux.

The data were calibrated on-site using the \textit{CLIC} package of the IRAM GILDAS data reduction software and typical flagging \citep{clic}. Images were deconvolved using the \textit{MAPPING} package on the calibrated visibility tables. The Clark cleaning algorithm was utilized to construct our clean map. Our data was processed with between 50-100 iterations, natural weighting for coverage points within our (u,v) grid, a default velocity bin width of 28 km s$^{-1}$, and an average cell size of 1.3$\arcsec$ for our 128$\times$128 image. For G20-2, G13-1, G14-1, \& G08-5, the emission line signal-to-noise ratio (SNR) was low enough ($<$5) to require re-binning in velocity space to 119 km s$^{-1}$. Typical RMS flux values within the resultant cubes were 2-3 mJy beam$^{-1}$ (exact flux errors in Table \ref{tab:observed}). The observations discussed above build upon a sample from a previous program \citep[X02C in the June-November 2013 period; see][]{fisher2014}.

Emission line analysis for our mm-wavelength observations was done using the Common Astronomy Software Applications (CASA)'s spectral line tool. For each target, we extracted spectra using a beam-size aperture (except for C13-1 and D15-3, which required larger apertures due to being marginally-resolved) and directly integrated over our CO[1-0] line to obtain an upper limit estimate for the velocity integrated flux ($F_{\rm CO}$; in Jy km s$^{-1}$). The CO emission lines for our two lowest-redshift targets (C13-1 \& D15-3; z$\sim$0.07) exhibit a `double-peaked' line-shape (see Fig. \ref{fig:co_spec}, row 3). For these two targets we have included position-velocity diagrams in Fig. \ref{fig:co_spec}. 

To estimate upper limits for the molecular gas masses and implied mass fractions ($f_{gas}$=$M_{\rm gas}/(M_{\rm gas} + M_{*}$)) we used the following expression (similar to \citealt{tacconi2013}):

\begin{equation}
M_{\rm gas} = \rm 1.75\times 10^{9} \left(\frac{\alpha_{CO}}{4.36}\right) F_{\rm CO}\ R_{\rm J1}\ \lambda_{\rm obs}^{2}\ D_{\rm L}^{2}\ (1+z)^{-3}
\end{equation}

\noindent where F$_{\rm CO}$ is the CO[1-0] flux in Jy $\kms$, R$_{\rm J1}$ is the transition coefficient (equal to 1 for the ground-state), $\lambda_{\rm obs}$ is the observed wavelength in mm, and D$_{\rm L}$ is the luminosity distance in Gpc. We note that this formula incorporates a correction for the 36\% increase due to the Helium fraction of molecular clouds. Similar to \cite{fisher2014}, we assume the following H$_{2}$ mass-to-CO luminosity ratio ($\alpha_{\rm CO}$; see \citealt{bolatto2013} for a thorough review of this ratio for all galaxy populations):

\begin{equation}
\rm \alpha_{CO} = M_{gas}/L'_{CO} \approx 3.1\ \Msun (K\ km\ s^{-1}\ pc^{2})^{-1}
\label{eq:aCO}
\end{equation}

\noindent for our six targets. While we acknowledge that choice of $\alpha_{\rm CO}$ is decisive in determining whether or not the galaxies appear gas rich, we point out that our assumed value is slightly conservative and defend this high redshift choice for DYNAMO systems using arguments presented in \S \ref{sec:2}. (As we will see in \S\ref{colddust}, our $T_{dust}$ estimates directly support this choice of $\alpha_{\rm CO}$.) The gas consumption timescale was estimated as $t_{\rm gas} = M_{\rm gas}/\rm SFR$. Observed fluxes and line widths can be found in Table \ref{tab:observed}; inferred values for $M_{gas}$ and $f_{gas}$, can be found in Table \ref{tab:inferred}. 

%-------------------------------------------------------
%              HERSCHEL IR/DUST OBSERVATIONS
%-------------------------------------------------------

\subsubsection{Herschel IR Observations}\label{sec:hers}
Four DYNAMO galaxies (D00-2, D13-5, D15-3, and G03-2) were observed in July 2012, using \textit{Herschel}'s PACS instrument in the 70 and 160$\mu m$ wavebands. The raw data was extracted from the \textit{Herschel} Science Archive and reduced \& analyzed using ESA's \textit{Herschel} Interactive Processing Environment (HIPE; \citealt{ott2010}) software and the current version of the reduction pipeline. Fluxes values were estimated using HIPE's \textit{annularAperturePhotometry} module coupled with the suggested aperture sizes stated in the HIPE data reduction manual\footnote{Wiki - http://herschel.esac.esa.int/hcss-doc-14.0/index.jsp}. Flux values were re-scaled via aperture corrections given in \cite{zoltan2014} to account for lost light due to a fixed aperture size. The quoted errors in flux were estimated using ESA's HIPE software which averages background noise levels within a similar region size, but offset 10$\arcsec$ from the target.

Five additional DYNAMO galaxies (C08-2, I09-1, C14-2, D14-1, and G14-1) were observed as part of the Herschel ATLAS (hereafter, H-ATLAS; \citealt{eales2010}) survey. H-ATLAS targets have observations with both PACS (100 \& 160$\mu m$) and SPIRE (250, 350, \& 500$\mu m$) cameras. For these galaxies fluxes have been extracted directly from the H-ATLAS catalog (\citealt{bourne2016}). Details on data reduction and flux estimation for these targets can be found in \cite{eales2010}.

\subsubsection{Single-T ``Greybody" Fitting}\label{sec:modBB}
Dust mass has been shown to be an excellent tracer of hydrogen gas mass \citep[reviewed in][]{bolatto2013}. %\cite{draine2007} modelled dust emission within galaxies and found that emission beyond 100$\mu$m (a similar wavelength regime) originates primarily from dust in thermal equilibrium with the interstellar radiation field and traces the vast majority (98-99\%) of the dust mass.
Dust emission in galaxies is commonly modeled via a single temperature modified blackbody (also known as a ``grey body") approximation. Modelling the dust in this manner makes two noteworthy assumptions: 1) that all dust grains share a common temperature and 2) the dust distribution is optically-thin. In recent years, spectral energy diagram (SED) fitting methods have been developed which fit more complex dust models - namely those in which the dust grains are represented by a range of temperatures \citep[e.g][]{draine2007}.

\begin{figure*}[!htb]
\begin{center}
\includegraphics[scale=0.45]{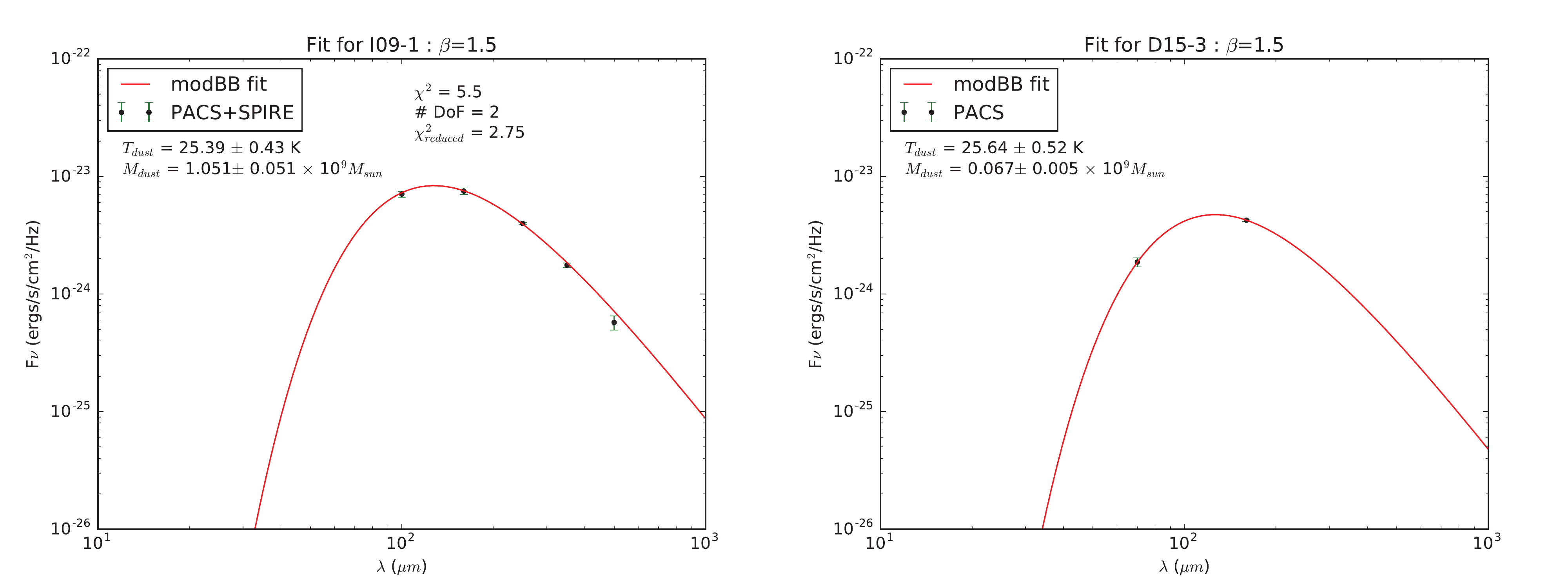}
\end{center}
\caption{For 9 of the 13 galaxies presented in this paper, dust masses obtained from fitting a modified blackbody model to the \textit{Herschel} waveband data were used to infer the gas mass. Here, we show example fits (I09-1 and D15-3) for which we have chosen $\beta$=1.5. For our sub-sample of DYNAMO galaxies observed with \textit{Herschel}, we consistently find $T_{dust}<$30K and dust masses which suggest $f_{gas}$$\sim$10-40\%.}\label{fig:modBB}
\end{figure*}

In this paper, we fit the following equation, representing a modified blackbody, to the \textit{Herschel} data:

\begin{equation}\label{eq:modBB}
f_{\nu} = \frac{M_{dust}}{D_{L}^{2}}\ \kappa_{abs}\ B_{\nu}(T_{\beta})
\end{equation}

\noindent where $D_{L}$ is the luminosity distance, $\kappa_{abs}$ is the emissivity (or the absorption coefficient, where $\kappa_{abs}=\kappa_{0}(\lambda_{0}/\lambda)^{\beta}$), $B_{\nu}$ is the Planck function for a single-temperature dust model, and $T_{\beta}$ \& $M_{dust}$ are left as fitted parameters. Surveys of the far-infrared \textit{Herschel} colors of galaxies have revealed that $1\leq \beta \leq2$ \citep{boselli2012, auld2013, bianchi2013}. For this work, we adopt a $\beta$=1.5 value consistent with that in \cite{draine2007}. Example fits for I09-1 and D15-3 are given in Fig.\ref{fig:modBB}.

In the paragraphs that follow, we outline the uncertainties introduced by estimating the dust mass (and, similarly, the gas mass) in this manner.

The largest source of uncertainty in the greybody dust model is the assumption of a single temperature for all dust emission. Galaxies likely have a range of dust temperatures \citep{draine2007}. Emission from $\sim 20-150 \mu m$ is dominated by warm dust mostly heated by young stars. Emission at longer wavelengths may be driven by lower energy photons, and may have lower temperatures. This introduces a systematic uncertainty in the mass estimates we derive from this technique as it does not explicitly account for contributions from a cold dust (10 - 15K) component. Recent comparisons of the full-SED and modified blackbody methods (e.g. \citealt{cortese2012,bianchi2013,berta2016}) reveal that the modified blackbody model systematically underestimates $M_{dust}$ by 20-50\% compared to \cite{draine2007} models. We therefore acknowledge that the  dust (and, similarly, gas) mass estimates for our \textit{Herschel} sample may be greater than those presented here. 

As we discuss above, not all galaxies in our {\em Herschel} sample have the same wavelength coverage for the IR SED. To account for this greybody fits for H-ATLAS targets were re-processed with solely the 100 \& 160$\mu$m fluxes and this resulted in estimates for $M_{dust}$ and $T_{dust}$ that were in agreement (to the 5-point fitted values) within $\sim$20\%.

Use of a Galactic dust-to-gas ratio has a number of limitations. Firstly, there is significant uncertainty in its value: observations of solar metallicity galaxies (such as DYNAMO) produce dust-to-gas ratios which exhibit a scatter of about 0.4 dex \citep{remyruyer2014}. Moreover, the mass inferred from dust conversion represents a galaxy's \textit{total} gas content, whereas CO emission specifically traces the molecular gas (see \S\ref{sec:co_obs}). To date, the H$_{2}$/HI ratio in gas rich disks remains poorly constrained and while typical values of H$_{2}$/HI in local spirals is about 1/3 (\citealt{saintonge2011b}), in high-$z$ disks it may be closer to parity (\citealt{obreschkow2009}). We also acknowledge that assumption of a universal dust-to-gas ratio across our sample is likely too simplistic and might bias our result. In light of such, we estimate that a more realistic uncertainty in our dust results is $\sim\pm$0.3 dex (around a factor of 2). 

Despite these caveats, \cite{genzel2015} find that, on average, CO and dust techniques provide consistent gas mass estimates of galaxies across a range of redshifts (see their Eq. 2). DYNAMO galaxies G14-1, D15-3, and D13-5 fall within both our IR and CO samples (D13-5 CO values from \citealt{fisher2014}) and estimates for their molecular gas masses using both techniques are in good agreement (within a few percent). Therefore, we adopt the same formula (e.g. Equation \ref{eq:aCO}) as well as a similar conversion factor (D:G=0.01) to convert our fitted dust values to gas masses. Final $M_{gas}$ estimates (with errors) and corresponding $f_{gas}$ values can be found in Table \ref{tab:inferred}.

\subsection{System kinematics and SFRs}
 For 9 of the 13 galaxies discussed in this paper, we utilize the kinematic modeling results of \citealt{bekiaris2016}. The authors refer readers interested in a more in-depth discussion on the methodology for determining kinematic properties of galaxies in DYNAMO to \citealt{bekiaris2016}. The global gas velocity dispersion ($\sigma$) and rotational velocity ($v_{c}$) values included in our analysis (listed in Table \ref{tab:inferred}) have been extracted directly from the Table C2 in \citet{bekiaris2016} (note: $v_{c}$ is taken to be V$_{2.2R}$). For one galaxy, G08-5, we use kinematic values obtained from disk modelling of high-resolution Gemini maps \citep{fisher2017}.
 
All of the DYNAMO galaxies in our sample have IR observations using the Wide-field Infrared Survey Explorer (mission paper by \citealt{wright2010}; data available at http://irsa.ipac.caltech.edu/Missions/wise.html). To estimate star formation rates, we use Equation 2 from \cite{lee2013} and flux values from WISE band-4. SFRs for our sample of galaxies are listed in Table \ref{tab:inferred}.

%-------------------------------------------------------
%                RESULTS
%-------------------------------------------------------
\section{Results \& Discussion}\label{sec:4}
Estimates for gas mass ($M_{gas}$) and gas mass fraction ($\fgas$) for the 12 detected targets in our sample (we provide an upper limit for G13-1) are found in Table \ref{tab:inferred}. We report average $\fgas$ values of 0.20 and 0.23 for the systems with PdBI CO[1-0] and \textit{Herschel} data, respectively. Consistent with this, we find $<f_{gas,\rm H-ATLAS}>=0.25$ and $<f_{gas,\rm PACS}>=0.22$. This suggests excellent agreement between the two estimation techniques and is consistent with previous work on the sample ($<f_{gas}>$=0.2, from \citealt{fisher2014}). In all cases, the gas mass estimates fall below the estimated dynamical mass values (derived using $v(r)$=V$_{2.2}$ and ~$r$=2.2 $\times$r$_{1/2,r}$ from \citealt{bekiaris2016}) and, in most cases, the implied gas mass comprises about 10-15\% of the system mass.

Of the 12 detected galaxies, D15-3 has the lowest reported gas fraction ($\sim$11\%) and G03-2 the highest ($\sim$44\%). The mean value for molecular gas fraction in blue-sequence $z=0$ galaxies is about 4\%. Total gas is a factor of a few larger following COLDGASS, see \citealt{saintonge2011b}. This shows that DYNAMO galaxies have substantially higher molecular gas content presumably fueling their higher rates of star formation.

%-------------------------------------------------------
%           Non-detection of CO[1-0] in G13-1
 We do not observe a significant emission source in the CO map for G13-1. We place upper limit constraints on its molecular gas mass. We measure an upper limit to CO(1-0) flux of 0.149~Jy~km~s$^{-1}$. Using the same $\alpha_{CO}$ as in Equation~1 implies an $\fgas$ $<$4\%. Thus, we find G13-1 to be comparatively gas-poor with respect to its fellow DYNAMO members, and more in line with what is routinely observed in local SFGs \citep{saintonge2011b}. Moreover, the lower CO flux per unit SFR of G13-1 is consistent with the interpretation of this galaxy as a merger based on both kinematic and HST morphology \citep{fisher2017}. Similarly, DYNAMO galaxy H10-2 was undetected by \cite{fisher2014} and subsequently determined with the same morphological and kinematic analysis to be more consistent with advanced stage merging galaxies.
%-------------------------------------------------------

In DYNAMO-I, \citeauthor{green2014} use the Kennicutt-Schmidt law, which defines the relationship between a galaxy's surface density of gas and star formation, to report estimates for the total (HI + H$_{2}$) gas content of the sample. The molecular gas fraction reported in this paper seem to suggest that the majority ($\sim$40-70\%) of the gas content is molecular. We interpret this as a likely consequence of the fact that DYNAMO observations of star forming regions probe within the inner part of the disk where (as seen in \citealt{leroy2008a}) the gas content is mainly H$_{2}$-dominated. Alternatively, the galaxies may be so gas enriched that the hydrogen gas over a larger fraction of the disk doesn't remain in the atomic phase, but instead transitions into the molecular state (due to increasing density); beyond $\Sigma_{gas}\sim$10 $\Msun$/pc$^{2}$ pressure allows HI to transition into H$_{2}$. Within DYNAMO, average gas densities are observed to be well above this threshold.

As stated in \S\ref{sec:3}, two of the lower redshift targets observed with PdBI were marginally-resolved. This allows for the construction of position-velocity diagrams (seen for C13-1 and D15-3 in Fig. \ref{fig:dispVSgasfrac}). The rotational velocities determined by \cite{green2014} are a factor of $\sim$1.5x higher than that suggested by the CO kinematics (even correcting for inclination); this is likely due to the fact that, again, the observed CO emission is probing only the inner part of the disk (\citealt{leroy2008a}).

%-------------------------------------------------------
%           COLD DUST WITHIN DYNAMO?
%-------------------------------------------------------

\begin{figure*}[!htb]
\begin{center}
\includegraphics[scale=1.0]{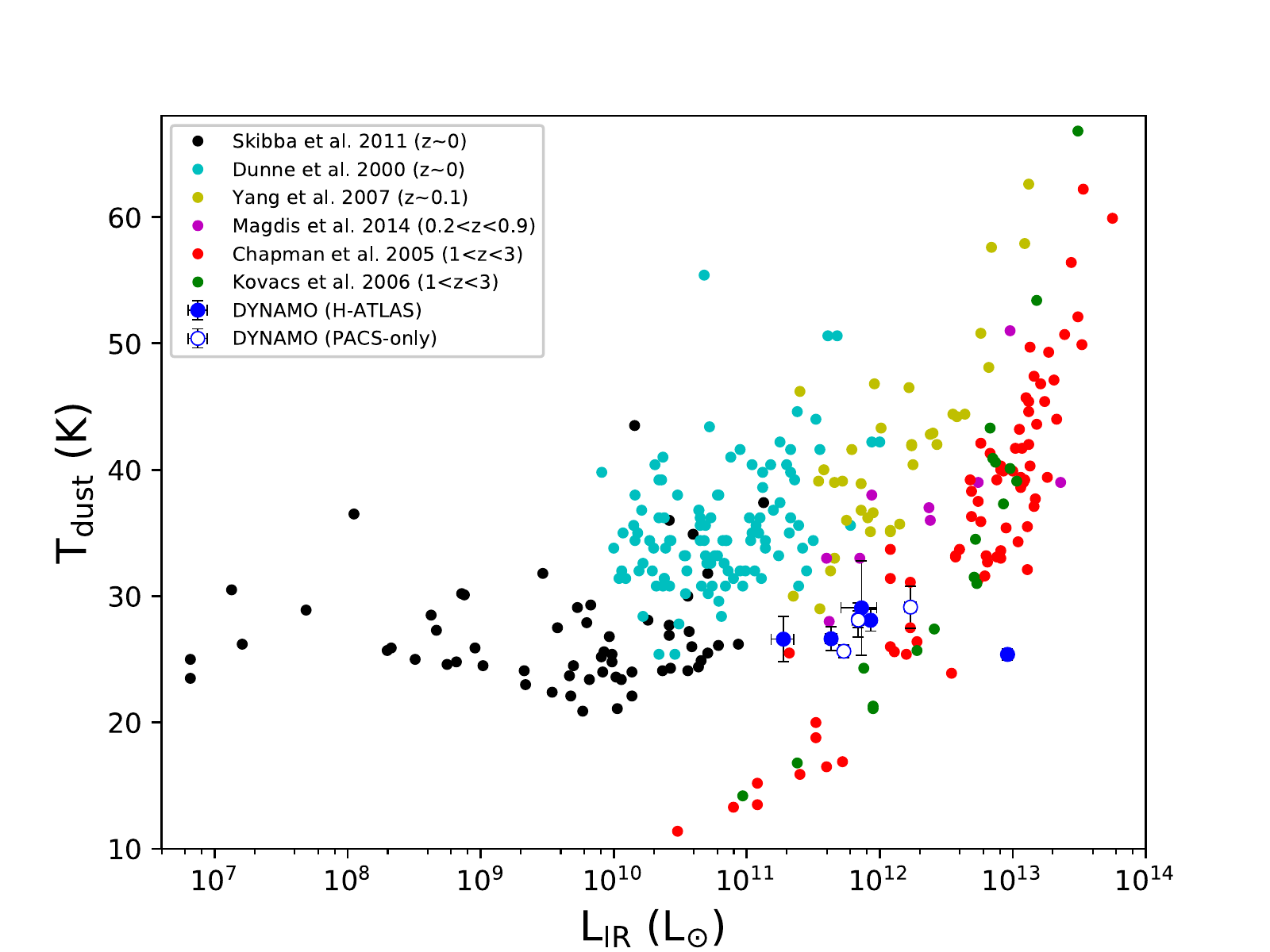}
\end{center}
\caption{DYNAMO galaxies (blue open and closed circles, with errors) are IR luminous, yet exhibit cold dust temperatures. We derived T$_{dust}$ values using the modified blackbody equation in \S\ref{sec:modBB} (Eq. \ref{eq:modBB}) and total IR luminosities were estimated from Herschel PACS+SPIRE fluxes (\citealt{galametz2013}; Eq. 3). Adapted from Fig. 3 in \cite{hwang2010}, we include galaxies from KINGFISH (black circles; \citealt{skibba2012}), IR-luminous galaxies at z$\sim$0 (cyan; from \citealt{dunne2010}), intermediate redshifts (\citealt{yang2007} in yellow, \citealt{magdis2014} in magenta) and sub-mm galaxies (SMGs; red, from \citealt{chapman2003} \& \citealt{kovacs2006}, in green) for reference.}
\label{fig:Tdust}
\end{figure*}

\subsection{Cold dust within DYNAMO galaxies?}\label{colddust}
The average $T_{dust}$ for DYNAMO galaxies measured in this paper is $27.5 \pm 0.52$ K. There is excellent agreement between the H-ATLAS and PACS-only samples: we find $<T_{dust}>=27.16 \pm 0.87$ K and $27.76 \pm 0.58$ K, respectively. The highest value for $T_{dust}$ in DYNAMO galaxies measured in this paper is $\sim$30 K.

We use the IR-band information to evaluate the total IR luminosity for DYNAMO systems (\citealt{galametz2013}, Eq. 3 \& Table 3) and in Fig. \ref{fig:Tdust}, we compare the dust temperature and luminosity of DYNAMO galaxies to other characteristic samples. Local spirals have similar dust temperatures as DYNAMO galaxies, with $<T_{dust}>$ $\sim$26.8 $\pm$ 0.71 K \citep{skibba2012}. However, these local spirals are typically 1-2 orders of magnitude fainter in IR luminosity. In the local Universe, systems with similar IR luminosities (and hence similar SFR) to that of DYNAMO galaxies have average dust temperatures that are 43.2 $\pm$ 2.1 K \citep{yang2007}. This is because those starbursts are mostly very concentrated, because they come from mergers, for example. This is considerably (about a factor of two) higher than that observed in DYNAMO galaxies. These results argue for the presence of cold dust within DYNAMO systems and suggest that a greater fraction of the interstellar dust is not being heated as it is in $z=0$ galaxies with similar star formation rates.

\cite{yang2007} suggest that the high dust temperatures observed in their sample is a consequence of the compact nature of the IR emission. Here we derive a simple and straightforward relationship between the dust temperature and fundamental galaxy parameters. 

We begin with the Stefan-Boltzmann Law:

\begin{equation}
F=\sigma_{b}\ T_{\rm eff}^{4} = \frac{L}{2\pi R_{d}^{2}}
\end{equation}

\noindent which can be solved for the effective temperature ($T_{\rm eff}$). This is then re-arranged to the following form:

\begin{equation}
T_{dust} \approx T_{\rm eff} = \left(\frac{L}{2\pi R_{d}^{2} \sigma_{b}}\right)^{1/4}
\end{equation}

\noindent where $\sigma_{b}$ is the Stefan-Boltzmann constant (in cgs units), $R_{d}$ is a characteristic radius for the disk, and $L$ is the dust luminosity (defined as $L= \rm SFR \times \epsilon c^{2}$, where $\epsilon=8\times$10$^{-3}$ for a Kroupa or Chabrier initial mass function). One then finds that a galaxy's dust temperature should roughly scale with its star formation rate and disk size:

\begin{equation}\label{eq:Tdust}
T_{dust} \propto \rm SFR^{1/4}\ R_{d}^{-1/2}.
\end{equation}

The above proportionality implies that the bulk of the dust in galaxies with colder dust temperatures is on average more distant from the radiation source. More generally, the dust temperature is a function of the local interstellar radiation field. If the star formation is distributed throughout these galaxies and the SFR density is lower then the temperatures will be colder. The dust in DYNAMO galaxies is then likely more extended than one might expect in a typical ULIRG at $z=0.1$ with the same SFR. DYNAMO systems are selected to be rotating (ie. disks), a criterion not imposed on the sample presented in \cite{yang2007}. Therefore, dust in DYNAMO systems are likely (on average) less compact and, consequently, less heated.

In local spirals (for example, KINGFISH survey galaxies) the Jeans length is comparatively small - thus, collapse occurs more readily and the dust is mostly located in regions where stars are actively forming. This is not assumed to be the case for the DYNAMO galaxies presented here, whose higher gas velocity dispersions predict larger Jeans lengths. %Low observed dust temperatures are also a likely consequence of the fact this population of galaxies is so gas rich: a larger fraction of the gas and its dust are simply not next to regions that are collapsing and forming stars. In reality, the cold dust temperatures we infer are most likely due to a combination of these factors (e.g. the galaxies are, indeed, more extended but they're also observed to be gas rich).

We conclude by noting that the state of the ISM of turbulent disks in the DYNAMO sample appears to be most similar to main-sequence galaxies at $z\sim 1-2$ \citep{chapman2003}, which we show in Fig. \ref{fig:Tdust}. We highlight that \cite{chapman2003} is the only sample that overlaps with DYNAMO galaxies in both L$_{\rm IR}$ and $T_{\rm dust}$. This is yet again an example of the similarity between DYNAMO galaxies and $z\sim$1-2 main-sequence disks.

Dust temperature is frequently used as an indicator of CO conversion factor (reviewed in \citealt{bolatto2015}). The dust temperatures we measure, therefore, also provide practical information about measuring gas mass via CO[1-0] flux on similar galaxies in DYNAMO. We find dust temperatures $\sim$27K for our DYNAMO-\textit{Herschel} sample. \cite{magnelli2012} suggest a critical value for $T_{\rm dust}$ of $\sim$30K (see also \citealt{solomon1997,tacconi2008}). For SFGs that are dominated by self-gravitating giant molecular clouds and which generally have colder dust temperatures ($<$30K) they prescribe a Milky Way-like value for $\alpha_{\rm CO}$. For systems observed with hotter dust ($>$30K) their results suggest $\alpha_{\rm CO}$ = 1\ M$_{\odot}\rm\ (K\ km\ s^{-1}\ pc^{2})^{-1}$ may be more appropriate. %A major benefit of using dust temperature to indicate an appropriate value for $\alpha_{CO}$ is that it is entirely independent of any ambiguity in kinematic classification. 
%In \S\ref{sec:2} we nonetheless present substantial evidence that our DYNAMO systems are unlikely to be mergers, and instead best resemble disks which lie close to the z$\sim$1.5 star forming main-sequence. This implies that the dominant mode of star formation in our DYNAMO galaxies is secular. 
We note that those few galaxies that have both IR and CO data measure very consistent gas masses through these independent methods (similar to results of \citealt{genzel2015} for high-z main-sequence galaxies). Moreover, \cite{fisher2014} finds a similar, Milky Way-like, $\alpha_{\rm CO}$ for DYNAMO galaxies using the stellar mass surface density \citep[as prescribed in][]{bolatto2013}. It therefore appears that all efforts to constrain CO-to-H$_{2}$ conversion factors in DYNAMO disks are consistent with our choice of $\alpha_{\rm CO}$ in \S\ref{sec:co_obs} (Equation \ref{eq:aCO}).

\begin{figure}[!htb]
\begin{center}
\includegraphics[scale=0.56]{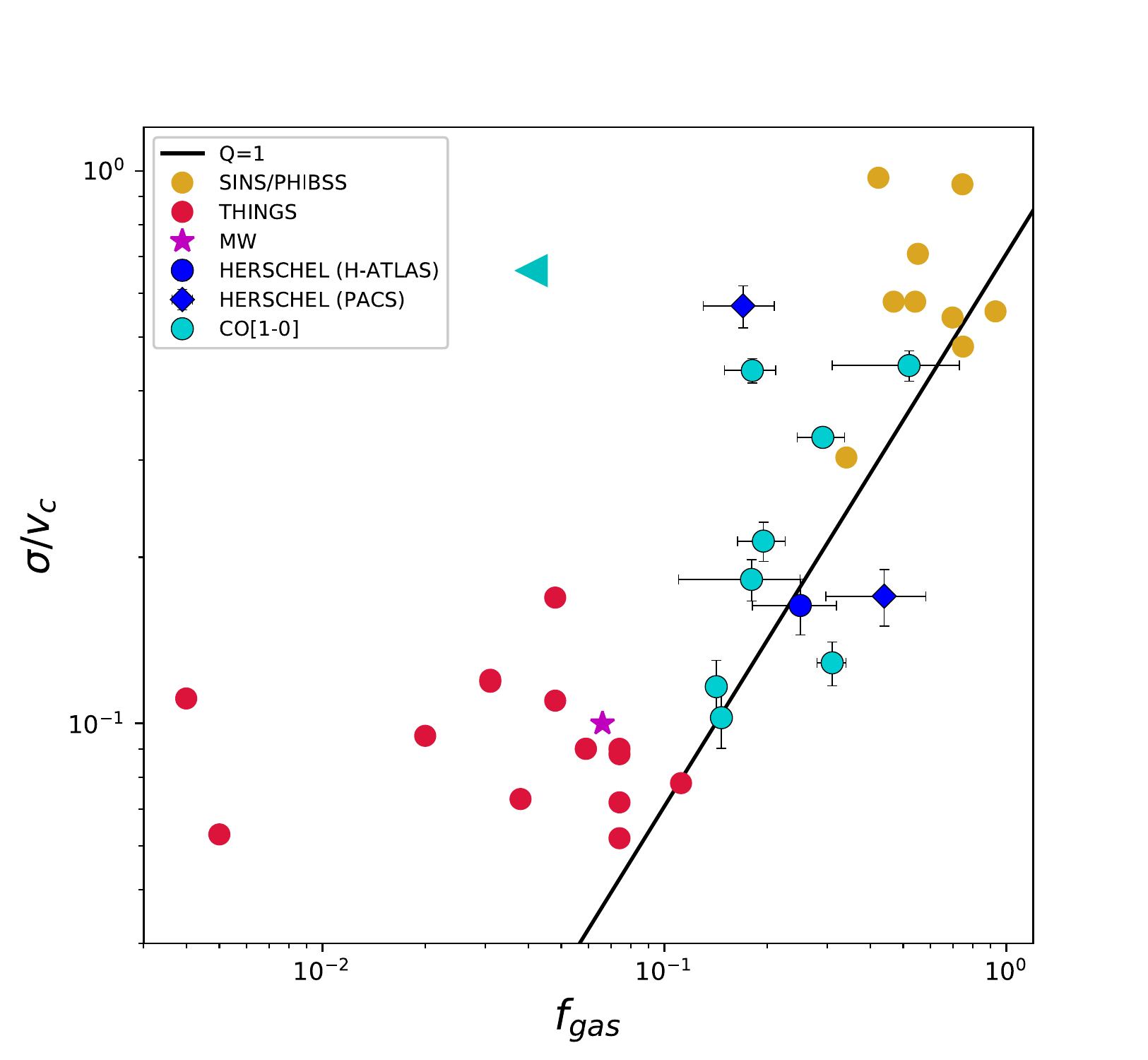}
\end{center}
\caption{The ratio between mean velocity dispersion to rotational velocity plotted against the molecular gas mass fraction ($f_{gas}$; using CO-derived values where possible).  We include the 3 detected systems from \cite{fisher2014}. To place these DYNAMO galaxies (blue and cyan, with upper limit value for G13-1 as leftward arrow) in context, we include nearby disk galaxies from the THINGS survey (in red; \citealt{walter2008,Ianjamasimanana2012}) and high-$z$ star forming galaxies observed as part of PHIBSS (in gold, kinematics from SINS; \citealt{genzel2013,fs2009}). The solid line corresponds to the relationship between $f_{gas}$ and $\sigma$/$v_{c}$ derived in the text, assuming instability theory (Eq. \ref{eq:result1}).}
\label{fig:dispVSgasfrac}
\end{figure}

%-------------------------------------------------------
%           sigma/V vs everything
%-------------------------------------------------------

\subsection{\texorpdfstring{$\sigma$/$v_{c}$ vs f$_{gas}$}{sigma/V vs fgas}}

We utilize the kinematic information from DYNAMO-I to investigate this relationship between the molecular gas content and dispersion within the disks in Fig. \ref{fig:dispVSgasfrac}. The DYNAMO sample in this figure is constructed from the 9 unique galaxies from this paper with kinematics and 3 detected systems from \cite{fisher2014}, utilizing CO-derived $f_{gas}$ values where possible. Nearby disk galaxies from The HI Nearby Galaxy Survey (THINGS, seen in red, where we have excluded dwarf galaxies and selected out systems with $M_{star} >$10$^{9} \Msun$; \citealt{walter2008,Ianjamasimanana2012}) and high-$z$ star forming galaxies observed as part of IRAM Plateau de Bure High-$z$ Blue Sequence Survey (PHIBSS; seen in green, with kinematics from the SINS survey; \citealt{tacconi2013,genzel2013,fs2009}) have been included to place the DYNAMO results (blue and cyan for Herschel and PdBI, respectively) in context. Note that we have added in a 10 km/s correction for thermal broadening to the sigma values for THINGS galaxies. The DYNAMO data fills in parameter space between the high $f_{gas}$ and $\sigma/v_{c}$ values reported in PHIBSS-SINS ($z>2$) galaxies and the low-$z$ THINGS galaxies. DYNAMO data therefore is necessary to identify identify the $f_{gas}$-$\sigma/v_{c}$ relationship as a one-to-one correlation as opposed to two distinct groupings in parameter space.

We find that the data set in in Fig.~5 has a Pearson's correlation coefficient of r=0.77 (note: in our fit, we exclude our upper-limit values and the two, low-$f_{gas}$ outliers in THINGS). The best fit relationship to the data in Fig.~\ref{fig:dispVSgasfrac} returns a sub-linear slope, however with very large scatter. We find $log(\sigma/v_{c})\propto 0.7 \pm 0.6 \times log(f_{gas}) - 0.07\pm 0.35$. We acknowledge that for a different choice of $\alpha_{\rm CO}$, this correlation would not hold.

The correlation we observe in Fig.~\ref{fig:dispVSgasfrac} has been assumed, or predicted, by a number of previous authors when invoking marginal stability models of disks \citep[e.g.][]{swinbank2012,genzel2013,glazebrook2013}. In this model the stability of rotationally-supported disks ($v_c/\sigma\geq$1) is represented by Toomre's Q parameter (\citealt{toomre1964}):

\begin{equation}
Q_{gas} = \frac{\sqrt{2} v_{c} \sigma}{\pi G r \Sigma_{gas}}
\end{equation}

\noindent where we have assumed a flat rotation curve. In this relation, $v_{c}$ is the circular velocity, $\Sigma_{gas}$ is the surface density of the gas at radius $r$ and defined as:

\begin{equation}
\Sigma_{gas}= \frac{M_{gas}(r)}{\pi r^{2}}
\end{equation}

\noindent and we have taken $\sigma$ to be the local vertical gas velocity dispersion. 

Using $v_{c}^{2}$=$\frac{GM_{tot}(r)}{r}$ and defining $f_{gas}$=$\frac{M_{gas}(r)}{M_{tot}(r)}$, then Toomre's $Q_{gas}$ can be expressed as:

\begin{equation}
Q_{gas} = \sqrt{2}\ \frac{\sigma}{v_{c}} \frac{1}{f_{gas}}.
\end{equation}

A direct relationship between $\sigma$/$v_{c}$ and $f_{gas}$ (we note that this relation has been previously predicted by \citealt{genzel2011} and \citealt{glazebrook2013}) emerges when we set $Q_{gas}$=1 (the instability condition):

\begin{equation}\label{eq:result1}
\frac{\sigma}{v_{c}} = \frac{f_{gas}}{\sqrt{2}}.
\end{equation}
%------------Why is the following paragraph commented out entirely?
% Overall, the predicted trend between $f_{gas}$ and $\sigma$/$v_{c}$ (plotted with our observational results in Fig. 4) is well-reflected in the derived quantities for the DYNAMO sample: on average, ($\sigma$/$v_{c}$)/$f_{gas}$ $\sim$1.6. While we do observe significant spread this is likely to be the consequence of two factors: 1) a hard-line assumption of $Q_{gas}$=1 as a condition for collapse is not entirely accurate and 2) observational uncertainty. In Fisher et al. 2016 \textit{submitted}, we find that, for DYNAMO galaxies, $Q_{gas}\sim0.5-1.5$. As observed in the figure, DYNAMO galaxies span the region between the populations of typical local spirals (represented by THINGS) and high-z turbulent disks (similarly, PHIBSS/SINS) but appear most closely matched to $z=1.5-2$ galaxies.
The consistency between gas fraction and kinematics (as given in Eq. 10) has important implications for disk properties and the role that this gas rich mode of star formation plays in defining galaxy structure. As the dynamical timescale within a typical disk is of order $t_{dyn}\sim$10 Myr, we can assume that the gas within the disk is in hydrostatic equilibrium (e.g. $\frac{dP}{dz} = -\rho g_{z}$, where P is the pressure and $g_{z}$ is the vertical component of gravity). Then, using Gauss' Law for the gravitational contribution due to the gas one can show that:

\begin{equation}
g_{z,gas}=2\pi G \Sigma_{g}
\end{equation}

\noindent where as the stars and dark matter contribute \citep{binneytremaine}:

\begin{equation}
g_{z,stars+DM}=\frac{v_{c}^{2}}{r}\frac{z}{r}(1-f_{gas}).
\end{equation}

If one assumes that the pressure is entirely the result of turbulent motions of the gas (e.g. P=$\rho\sigma^{2}$), which extend above some scale height z=H, then:

\begin{equation}
\frac{dP}{dz} \approx \frac{\rho\sigma^{2}}{H}.
\end{equation}

\noindent Setting Eq. 10 equal to the condition for hydrostatic equilibrium (where $g_{z}$=$g_{z,gas}$+$g_{z,stars+DM}$) one finds that:

\begin{equation}
\frac{\sigma}{v_{c}} \approx \frac{H}{r}.
\end{equation}

\noindent This result can be inserted into Eq. \ref{eq:result1} ($Q_{gas}$=1) to find:

\begin{equation}
f_{gas} = \frac{\sqrt{2}H}{r}.
\end{equation}

This presents an important, physical explanation for the results presented in Fig. \ref{fig:dispVSgasfrac}: in a marginally stable disk, higher gas fractions naturally lead to to thicker disks. This is consistent with \citealt{glazebrook2013} where they show (via similar arguments) that the disk thickness is of order the Jeans length (this is also predicted in simulations; see \citealt{bournaud2009}). Moreover, \cite{bassett2014} find that stellar velocity dispersions in DYNAMO disks are high, implying that turbulent motions build thick disks. 

%-------------------------------------------------------
%                    SIGMA vs TDEP
%-------------------------------------------------------

\begin{figure*}[!htb]
\begin{center}
\includegraphics[scale=0.25]{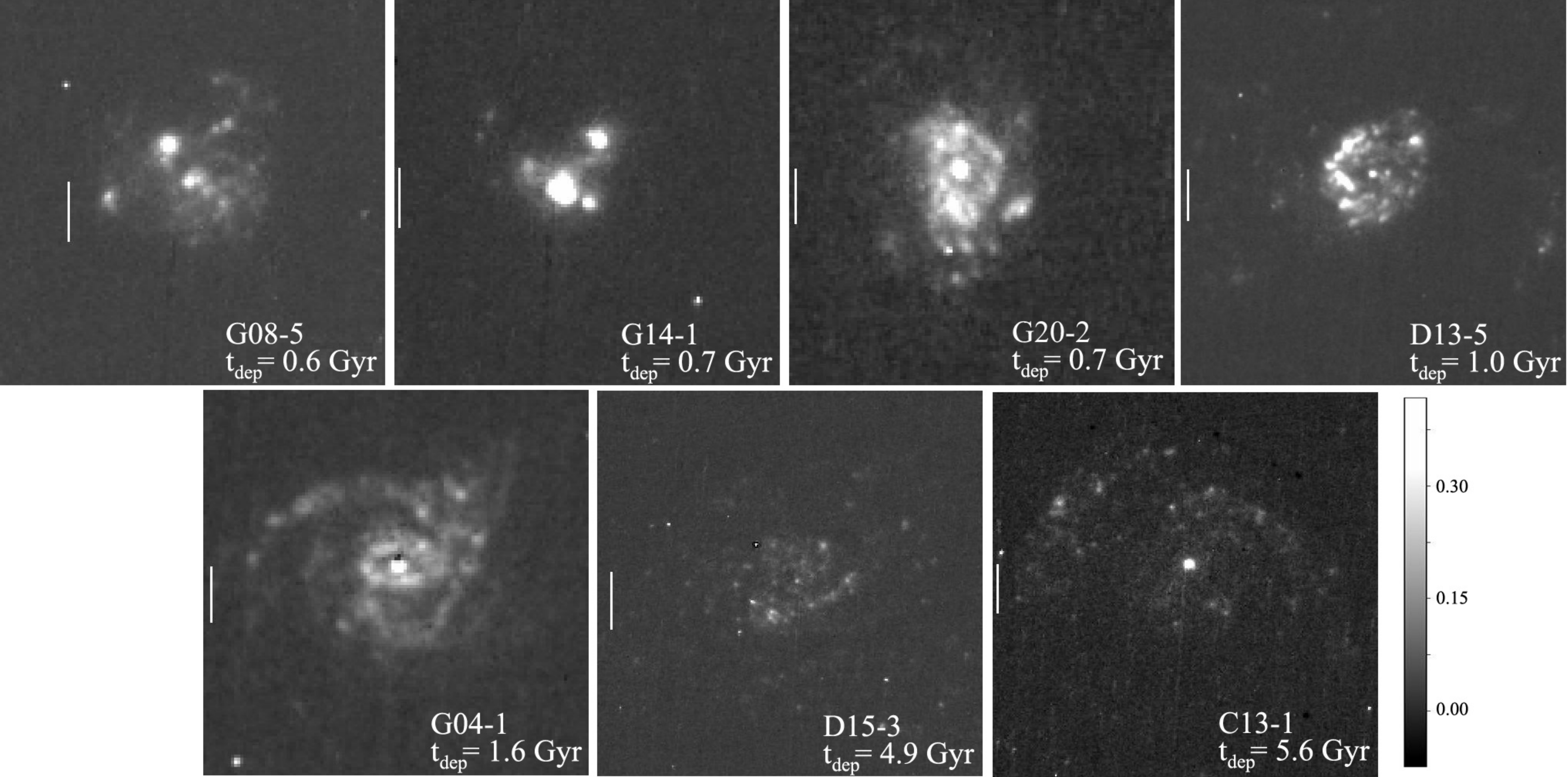}
\end{center}
\caption{High-resolution H$\alpha$ mapping (obtained using HST's Wide Field Camera on the Advanced Camera for Surveys (WFC/ACS); reproduced from \citealt{fisher2017}) of seven galaxies in DYNAMO. Galaxy maps are presented in order of increasing inferred depletion time. In this figure, we illustrate a developing trend evidenced in DYNAMO: galaxies with the shortest depletion times exhibit more prominent star forming clumps within their disks. The vertical, white bars in each of the seven panels above, represent $\sim$1 kpc on-sky and have been included for scale.}
\label{fig:tdepclumpmaps}
\end{figure*}

\subsection{Depletion times for DYNAMO galaxies}\label{tdep}
In the local universe there is a mild variation in global depletion times ($t_{dep}=M_{mol}$/SFR) on galaxy mass \citep{saintonge2011b}. At the $\sim$1 kpc scale, depletion times are found to be roughly constant in disk galaxies ($\sim$1 Gyr; \citealt{leroy2008a,bigiel2008,rahman2012}). At high redshift this value is more uncertain. For main-sequence systems around $z\sim$1.5 - 2 the average depletion time is a factor of 3-5 below the local spiral value \citep{daddi2010,tacconi2013}. In this paper we find DYNAMO galaxies have a range of depletion times from $\sim0.5-5$~Gyr. In Table \ref{tab:inferred}, we list the depletion times for our targets. 

% In Fig. \ref{fig:dispVSgasfrac}, we show that the subset of DYNAMO galaxies presented in this paper span the region between local spirals and the disks observed at 1.5$\leq z \leq$2. Indeed, the overall DYNAMO sample \citep{green2014} includes both turbulent disks and galaxies that better resemble $z\sim0$ star forming spirals.  Thus, we observe some of our disks to have longer depletion times (e.g. D15-3 and C13-1) while others are more similar to the systems in PHIBSS (e.g. G08-5, D14-1, G14-1). Note that due to assumptions about CO(3-2)/CO(1-0) (see \citealt{bolatto2015}), the estimated depletion times of galaxies in PHIBSS are uncertain by a factor of few. We can therefore use the \textit{range} of galaxies to track how gas depletion time evolves as the prominence of star forming clumps varies.

Galaxies G20-2, G14-1, and G08-5 from our sample all have depletion times below 1 Gyr (see Table \ref{tab:inferred}; D00-1 is consistent with 1 Gyr). These galaxies all have significantly higher $\sigma$/$v_{c}$ ratios (respectively $\sigma$/$v_{c} = 0.21, 0.44$, \& $0.33$) than both the rest of the DYNAMO sample and the THINGS sample. \cite{guo2015} argues that the ``clumpiness" of a galaxy is best defined by the maximum clump flux normalized by the galaxy flux observed within a target. Using this metric, we find that G20-2, G14-1, G08-5 have $L_{clump,max}/L_{total} = 0.1,0.3,$ and $0.1$ respectively \citep{fisher2017}.

Conversely, we observe longer depletion times for galaxies C13-1, D15-3 and D13-5 ($t_{dep}>$1 Gyr). These three longer depletion time systems have both lower values of $\sigma$/$v_{c}$ (0.12, 0.10, and 0.18, respectively), and significantly less prominent clumps ($L_{clump,max}/L_{total} = 0.02,0.03,$ and $0.08$). 

In Fig. \ref{fig:tdepclumpmaps}, we provide high-resolution HST H$\alpha$ maps for seven DYNAMO galaxies (G08-5, G14-1, G20-2, D13-5, G04-1, D15-3, and C13-1) with inferred depletion times to further illustrate this emerging trend between depletion time and clump prominence.

Our data thus far is consistent with lower gas depletion time galaxies having larger, more pronounced clumps and more turbulent gas (as indicated by higher ratios of $\sigma$/$v_{c}$). Nonetheless, this result should be taken as merely suggestive. In an forthcoming paper we intend to investigate this possible trend with a larger data set of NOEMA observations and ionized gas maps (Fisher et al {\em in prep}).

\section{Summary}\label{sec:5}
In this paper, we present gas mass estimates for a unique sample of 13 local galaxies ($z\sim$0.07 \& 0.1) whose kinematic and star forming properties closely resemble that observed in $z\approx 1.5$ star forming disks.
\begin{itemize}
    \item Six DYNAMO galaxies have been observed with the Plateau de Bure interferometer and targeting the CO[1-0] transition line (Fig. 1). Five are well-detected at $>$8$\sigma$ yielding CO fluxes and line luminosities consistent with gas mass fractions up to $\sim$30\%, assuming $\alpha_{\rm CO}$=3.1 $\Msun$(K km s$^{-1}$ pc$^{2}$)$^{-1}$.
    \item Fitting a modified blackbody function to existing Herschel IR observations (from PACS+SPIRE, Fig. 2) for 9 additional galaxies, we find that the dust within these galaxies is substantial ($M_{dust}$; 1-3$\times 10^{9}\Msun$) and cold ($T_{dust}<$30K, Fig. 3). Using a locally-derived dust-to-gas ratio (D:G$\sim$0.01) our fitted $M_{dust}$ predict high gas masses ($f_{gas}\sim$10-40\%) for our Herschel sample.
    \item We confirm the gas-rich nature of DYNAMO galaxies with a sample that is $\sim$5x larger than previous work and via multiple methods to reduce observational bias.
    \item Coupling the gas mass fractions with existing high-resolution kinematics in DYNAMO we report a linear relationship between $f_{gas}$ and $\sigma$/$v_{c}$ (see Fig. 4). Predicted from Toomre instability theory, this provides direct observational evidence of the role performed by the internal velocity dispersion of the gas in the formation of massive star forming clumps within galaxies. 
    \item We find that DYNAMO systems with depletion times most consistent with that of high-$z$ star forming disks ($t_{dep}$ $\sim$0.5 Gyr) also exhibit the highest ratios of $\sigma$/$v_{c}$ and (when imaging is available) very prominent clumps within their disks.
\end{itemize}

\section*{Acknowledgements}
The science done above is based on data obtained at the Plateau de Bure millimetre interferometer, which is operated by the Institute for Radio Astronomy in the Millimetre Range (IRAM), which is funded by a partnership of INSU/CNRS (France), MPG (Germany), and IGN (Spain) and the Australian Astronomical Observatory and the Australian National University's 2.3 meter telescope. We also include observations performed with the ESA \textit{Herschel} Space Observatory \citep{pilbratt2010}, in particular employing Herschel's large telescope and powerful science payload to do photometry using the PACS \citep{poglitsch2010} and SPIRE \citep{griffin2010} instruments.

HAW and RGA thank NSERC and the Dunlap Institute for Astronomy and Astrophysics for financial support. The Dunlap Institute is funded through an endowment established by the David Dunlap family and the University of Toronto.

DBF and KG acknowledge support from Australian Research Council (ARC) Discovery Program (DP) grant DP130101460. Support for this project is provided in part by the Victorian Department of State Development, Business and Innovation through the Victorian International Research Scholarship (VIRS).

%-------------------------------------------------------
%                       BIBLIOGRAPHY
%-------------------------------------------------------

\bibliographystyle{apj}
\bibliography{dynamo}

%-------------------------------------------------------
%                      TABLES
%-------------------------------------------------------
\clearpage
\clearpage
\begin{turnpage}
\begin{deluxetable}{lccccccccccc}
\tabletypesize{\scriptsize}
\tablecaption{IR and Radio Observations of DYNAMO Galaxies\label{tab:observed}}
\tablewidth{0pt}
\tablehead{\textit{Herschel}\\ 
PACS+SPIRE
\\
\colhead{Galaxy} &
\colhead{70$\mu m$} &
\colhead{100$\mu m$} &
\colhead{160$\mu m$} &
\colhead{250$\mu m$} &
\colhead{350$\mu m$} &
\colhead{500$\mu m$} &
\colhead{T$_{dust}$ (K)} &
\colhead{M$_{dust}$}\\
}
\startdata
C08-2 & - &	0.393$\pm$0.043 &	0.356$\pm$0.045 &	0.188$\pm$0.006 &	0.082$\pm$0.007 &	0.023$\pm$0.008 &	26.63$\pm$0.93 &	3.82$\pm$0.37\\
I09-1 & - &	0.707$\pm$0.041 &	0.75$\pm$0.048 &	0.398$\pm$0.006 &	0.176$\pm$0.007 &	0.057$\pm$0.008 &	25.39$\pm$0.43 &	105.12$\pm$5.05\\
C14-2 & - &	0.18$\pm$0.041 &	0.26$\pm$0.048 &	0.095$\pm$0.007 &	0.038$\pm$0.008 &	0.015$\pm$0.008 &	26.6$\pm$1.79 &	1.83$\pm$0.36\\
D14-1 & - &	0.484$\pm$0.041 &	0.466$\pm$0.048 &	0.198$\pm$0.007 &	0.081$\pm$0.008 &	0.044$\pm$0.009 &	28.09$\pm$0.86 &	5.89$\pm$0.52\\
G14-1 &	- &	0.119$\pm$0.041 &	0.112$\pm$0.048 &	0.047$\pm$0.007 &	0.021$\pm$0.008 &	-0.005$\pm$0.009 &	29.07$\pm$3.74 &	4.39$\pm$1.65\\
D00-2 &	0.195$\pm$-0.001 &	- &	0.299$\pm$0.027 &	- &	- &	- &	28.18$\pm$0.64 &	5.08$\pm$0.84\\
D13-5 & 0.598$\pm$0.061 &	- &	0.803$\pm$0.154 &	- &	- &	- &	29.13$\pm$1.66 &	10.44$\pm$3.81\\
D15-3 & 0.187$\pm$0.016 &	- &	0.424$\pm$0.009 &	- &	- &	- &	25.64$\pm$0.52 &	6.68$\pm$0.55\\
G03-2 &	0.072$\pm$0.012 &	- &	0.111$\pm$0.011 &	- &	- &	- &	28.11$\pm$1.36 &	5.16$\pm$1.18\\
\hline
\\
PdBI CO[1-0]
\\
&   
\colhead{t$_{int}$ (hr)} &
\colhead{A$_{beam}$} &
\colhead{$\Delta$v} (km s$^{-1}$) &
\colhead{F$_{CO}^{c}$} &
\colhead{L`$_{CO}^{d}$} &
\colhead{M$_{gas}^{a}$} \\
\\
\hline
\\
C13-1 & 1.05 & 5.94$\arcsec$$\times$4.19$\arcsec$ & 196.4 & 5.84$\pm$0.15 & 1.91$\pm$0.05 & 0.59$\pm$0.02\\
G20-2 & 0.90 & 9.46$\arcsec$$\times$4.71$\arcsec$ & 237.4 & 1.57$\pm$0.18 & 1.68$\pm$0.19 & 0.52$\pm$0.06\\
G13-1 & 1.99 & 5.94$\arcsec$$\times$4.64$\arcsec$ & - & 0.149$^{\dagger}$ & 0.155$^{\dagger}$ & 0.048$^{\dagger}$\\
G14-1 & 1.09 & 7.13$\arcsec$$\times$4.66$\arcsec$ & 235.6 & 1.69$\pm$0.20 & 1.594$\pm$0.19 & 0.49$\pm$0.06\\
G08-5 & 1.05 & 5.36$\arcsec$$\times$4.47$\arcsec$ & 353.3 & 2.44$\pm$0.27 & 2.29$\pm$0.26 & 0.71$\pm$0.08\\
D15-3 & 1.46 & 6.26$\arcsec$$\times$4.44$\arcsec$ & 360.8 & 12.8$\pm$0.25 & 3.02$\pm$0.06 & 0.94$\pm$0.02\\
\enddata
\tablenotetext{a}{Dust masses are in units of 10$^{7}$. All other masses are stated in units of 10$^{10}$ \Msun.}
\tablenotetext{b}{All Herschel waveband fluxes in Jy.}
\tablenotetext{c}{Integrated flux density (F$_{CO}$) in units of Jy km s$^{-1}$.}
\tablenotetext{d}{CO line luminosity in units of 10$^{9}$ K km s$^{-1}$ pc$^{2}$.}
\tablenotetext{$\dagger$}{Denotes quantities derived from 2$\sigma$ upper limit flux values.}
\end{deluxetable}
\end{turnpage}
\global\pdfpageattr\expandafter{\the\pdfpageattr/Rotate 90}

\clearpage
\clearpage
\begin{turnpage}
\begin{deluxetable}{cccccccccc}
\tabletypesize{\scriptsize}
\tablecaption{Properties of DYNAMO Galaxies\label{tab:inferred}}
\tablewidth{0pt}
\tablehead{
\\
\colhead{Galaxy} &
\colhead{$M_{*}$} &
\colhead{$z$} &
\colhead{SFR$_{\rm WISE}$} &
\colhead{$\sigma$/$v_{c}$} &
\colhead{Method} &
\colhead{$M_{gas}$} &
\colhead{$f_{gas}$} &
\colhead{$M_{dyn}$} &
\colhead{$t_{dep}$}\\
\colhead{ } &
\colhead{(10$^{10}$\Msun)} &
\colhead{ } &
\colhead{(\Msun yr$^{-1}$)} &
\colhead{ } &
\colhead{ } &
\colhead{(10$^{9}$\Msun)} &
\colhead{ } &
\colhead{(10$^{9}$\Msun)} &
\colhead{(Gyr)}
\\
}
\startdata
C08-2 & 1.09 & 0.0581 &    1.86 & - & Dust &  3.82$\pm$0.37 &	0.26 & - &	2.05$\pm$0.72\\
I09-1 & 25.2 &  0.1818 &    15.6 &	- & Dust &	105$\pm$5.1 &	0.29 &  - &	6.72$\pm$1.99\\
C14-2 & 0.56 &  0.0562 &    1.12 &	0.16$\pm$0.02 &	Dust &	1.83$\pm$0.36 &	0.25 &  34.91 &	1.63$\pm$0.52\\
D14-1 & 2.04 &  0.0736 &    4.12 & - &	Dust &	5.89$\pm$0.52 &	0.22 &  - &	1.43$\pm$0.66\\
G14-1 & 2.23 &  0.1324 &    6.90 & 0.44$\pm$0.02 &	Dust &	4.39$\pm$1.65 &	0.16 &  27.63 &	0.64$\pm$0.32\\
D00-2 & 2.43 &  0.0813 &	5.14 &	0.57$\pm$0.05 &	Dust &	5.08$\pm$0.84 &	0.17 &  5.33 &	0.99$\pm$0.37\\
D13-5 & 5.38 &  0.0753 &    6.27 &	0.18$\pm$0.02 &	Dust &	10.4$\pm$3.80 &	0.16 &  82.95 &	1.66$\pm$1.02\\
D15-3 & 5.42 &  0.0671 &    1.91 &	0.10$\pm$0.01 &	Dust &	6.69$\pm$0.55 &	0.11 &  143.1 &	3.50$\pm$1.75\\
G03-2 & 0.65 &  0.1295 &    4.60 &	0.17$\pm$0.02 &	Dust &	5.16$\pm$1.18 &	0.44 &  51.15 &	1.12$\pm$0.30\\
\\
C13-1 & 3.58 & 0.0788 & 1.06 & 0.12$\pm$0.01 & CO[1-0] & 5.91$\pm$0.15 & 0.14 &  134.7 &	5.58$\pm$1.81\\
G20-2 & 2.16 & 0.1411 & 7.79 & 0.21$\pm$0.02 & CO[1-0] & 5.22$\pm$0.59 & 0.20 &  30.61 &	0.67$\pm$0.25\\
G13-1 & 1.11 & 0.1388 & 12.2 & 0.66$\pm$0.03 & CO[1-0] & 0.48$^{\dagger}$ & 0.04$^{\dagger}$ &  17.36 &	-\\
G14-1 & 2.23 & 0.1323 & 6.90 & 0.44$\pm$0.02 & CO[1-0] & 4.94$\pm$0.59 & 0.18 &  27.63 &	0.72$\pm$0.25\\
G08-5 & 1.73 & 0.1322 & 12.2 & 0.33 & CO[1-0] & 7.11$\pm$0.79 & 0.30 &  49.53 &	0.57$\pm$0.21\\
D15-3 & 5.42 & 0.0671 & 1.91 & 0.10$\pm$0.01 & CO[1-0] & 9.36$\pm$0.18 & 0.15 &  143.1 &	4.89$\pm$2.42\\
\\
D13-5$^{\star}$ & 5.38  & 0.0753  & 6.27 & 0.18$\pm$0.02 & CO[1-0] & 11.86$\pm$0.36 & 0.18 &  82.95 &	1.89$\pm$0.94\\
G04-1$^{\star}$ & 6.47 & 0.1298 & 15.0 & 0.13$\pm$0.01 & CO[1-0] & 28.99$\pm$2.1 & 0.31 &  110.5 &	1.94$\pm$0.72\\
G10-1$^{\star}$ & 1.22 & 0.1437 & 15.7 & 0.44$\pm$0.03 & CO[1-0] & 13.45$\pm$2.15& 0.52 &  17.50 &	0.86$\pm$0.15\\
\enddata
\tablenotetext{$\dagger$}{Denotes quantities derived from 2$\sigma$ upper limit flux values. 
}
\tablenotetext{$\star$}{Values taken from \cite{fisher2014}. Note: G10-1 does not have an entry in the WISE catalog. For this reason, we use the H$\alpha$-derived SFR quoted in \cite{green2014}.
}
\end{deluxetable}
\end{turnpage}
\global\pdfpageattr\expandafter{\the\pdfpageattr/Rotate 90}

\end{document}